\documentclass[journal]{IEEEtran}

\ifCLASSINFOpdf

\else

\fi

\usepackage{amssymb}
\usepackage{graphicx}

\begin{document}

\title{Iterative Detection with Soft Decision in Spectrally Efficient FDM Systems}

\author{ Seyed Javad Heydari, Mahmoud Ferdosizade Naeiny, Farokh Marvasti}
\thanks{M. Ferdosizade Naeiny is with Department of Electrical Engineering,
Shahed University of Technology and S.J. Heydari and F. Marvasti
are with Advanced Communication research Institute (ACRI) and the
Department of Electrical Engineering, Sharif University of
Technology, Tehran, Iran, e-mails: m.ferdosizade@shahed.ac.ir,}

\maketitle

\begin{abstract}
In Spectrally Efficient Frequency Division Multiplexing (SEFDM)
systems the input data stream is divided into several adjacent
subchannels where the distance of the subchannels is less than
that of Orthogonal Frequency Division Multiplexing (OFDM) systems.
Since the subcarriers are not orthogonal in SEFDM systems, they
lead to interference at the receiver side. In this paper, an
iterative method is proposed for interference compensation for
SEFDM systems. In this method a soft mapping technique is used after
each iteration block to improve its performance. The performance of the
proposed method is comparable to that of Sphere Detection (SD)
which is a nearly optimal detection method.
\end{abstract}


\IEEEpeerreviewmaketitle

\section{Introduction}
In recent years, Orthogonal Frequency Division Multiplexing (OFDM)
has attracted significant research interest as a potential
solution for high-rate data service demands over wireless channels
\cite{SamTal02}, \cite{RapAnn02}. In OFDM systems, the total bandwidth
is divided into several orthogonal narrowband subcarriers. Since the
subcarriers have a $50$ percent overlapping, OFDM system leads to a
high spectrum efficiency comparing to the ordinary Frequency Division
Multiple Access (FDMA) systems. The lower data rate in each subcarrier
makes the OFDM systems have a great advantage to combat Inter-Symbol
Interference (ISI) and enable high degree of flexibility of resource
allocation among users. However, demands for broadband wireless
applications have grown significantly in recent years and a reliable transmission and
reception of high data rate information over a lower bandwidth has drawn
a lot of attention. Rodrigues and Darwazeh in \cite{RodDar02} and Xiong
in \cite{Xio03} introduced a multicarrier system that occupies half of the
bandwidth of an equivalent OFDM, but the detection was possible for real
alphabets (e.g. BPSK) only. Later, Rodrigues and Darwazeh in \cite{RodDar03}
and Hamamura and Tachikawa afterwards in \cite{HamTac04} proposed spectrally
efficient FDM systems in which subcarrier frequency separation is smaller than
the inverse of the FDM signaling period, so they deliberately violated the
orthogonality principle between carriers. Consequently, these systems have
considerable bandwidth efficiency in comparison to that of OFDM systems.

Rusek and Anderson have shown in \cite{RusAnd05} that the concept of
reducing the frequency separation between orthogonal subcarriers is a
dual case of the Mazo's time-domain transmission technique. Mazo aims to
transmit information faster than the Nyquist rate \cite{MazLan88,Maz75}.
In addition, they have proved that in presence of AWGN, the frequency
separation of the subcarriers can be reduced up to a limit, which is the
dual of the so-called Mazo's limit. As a consequence, if this limitation
is taken into account, no degradation is expected in the performance of
spectrally efficient FDM systems in comparison to OFDM systems.
Notwithstanding, the reliable detection of the information of such FDM
systems is still a very challenging issue since the optimal maximum
likelihood detection is very complex. In \cite{RusAnd05}, it has been
shown that it is safe to increase the signaling rate by $20\%$ without
expecting any performance degradation.

Transmission of the SEFDM signal can be performed with the
standard Inverse Fast Fourier Transform (IFFT)
\cite{AhmDar09,IsaDar10}. However, detection of SEFDM signals is
challenged by the need to extract the original signal from the
inter-carrier interference. The optimal detection of the SEFDM
signal requires brute force ML which can become extremely complex
\cite{RodDar03}. On the other hand, using linear detection
techniques such as Zero-Forcing (ZF) and Minimum Mean Square Error
(MMSE) constrains the size of the SEFDM system and the level of
bandwidth savings due to the ill-conditioning of the system caused
by the orthogonality collapse \cite{IsaChor08}. The Truncated
Singular Value Decomposition (TSVD) was proposed as an efficient
tool to overcome this deficit \cite{IsaKan2011}, but its
performance was far from Optimum detector. Therefore, Sphere
Decoder (SD) was proposed in \cite{KanChoRod09} to provide ML
performance at a much-reduced complexity. Nevertheless, the SD
complexity varies greatly depending on the noise. Convex
Optimization tools are also used to achieve a reliable detector
\cite{KanChoRodJapan09}, but this method suffers from the same
problems as the linear ones. In \cite{KanCho09}, quasi-optimal detector
combining Semi-definite Programming and SD was proposed; however,
the complexity still remains variable. In \cite{IsaDar11}, authors
proposed the use of the Fixed-complexity Sphere Decoder (FSD)
algorithm for the detection of SEFDM signal. The FSD algorithm could
no longer guarantee an optimum solution; However, its
sub-optimality may be traded-off with complexity. A Precoding
strategy that greatly simplifies the detection of the signal was
proposed in \cite{Precoding}. The strategy facilitates simpler
detection for the same bandwidth savings as an equivalent uncoded
SEFDM system. However, it also suffers from ill-conditioning of
the system.

In this paper, an iterative method is proposed  whose
performance is comparable with that of SD method with lower
complexity. This method is a modified version of the general
iterative method introduced in \cite{Mar01}. In the modified
version, a soft mapping operation is used in order to decide some symbols
after each iteration. It is proposed that the soft mapping
parameter is adaptively changed after each iteration block. In section
II the system model of SEFDM systems is discussed and in section
III the proposed iterative method is introduced. Then, in section
IV the performance of the iterative method is evaluated by
simulations.
\section{System Model}
In the SEFDM system, the input stream of the complex symbols is
divided into $N$ parallel low data rate streams. The generated
symbol streams are modulated by different carrier frequencies. If
$S(n)$ is the complex symbol of the $n$th data stream, then the
time domain baseband signal can be written as
\begin{equation}
\label{eq:SEFDMtime1}
x(t)=\frac{1}{\sqrt{N}}\sum_{n=0}^{N-1}S(n)e^{j2\pi n\Delta ft},\quad 0\leq t< T,
\end{equation}
where $\Delta f$ is the spacing between the adjacent carrier
frequencies. The main difference between SEFDM and OFDM systems is that
the distance of carrier frequencies in SEFDM systems is only a
fraction of the inverse of the FDM symbol period $T$, i.e.
\begin{equation}
\label{eq:freqdist}
\Delta f=f_k-f_{k-1}=\frac{\alpha}{T},\quad \alpha<1
\end{equation}
As a result, the required bandwidth is reduced by a factor of
$1-\alpha$, at the expense of the loss of orthogonality between
carriers. The time domain samples of SEFDM signal are as follows:
\begin{equation}
\label{eq:SEFDMtime21}
X(k)=x(\frac{kT}{N})=\frac{1}{\sqrt{N}}\sum_{n=0}^{N-1}S(n)e^{j\frac{2\pi\alpha
nk}{N}},\quad 0\leq k< N-1,
\end{equation}
If $\mathbf{X}=[X(0),X(1),...,X(N-1)]^T$ and
$\mathbf{S}=[S(0),S(1),...,S(N-1)]^T$ then the above equation can
be written in the matrix form of
$\mathbf{X}=\mathbf{F}\mathbf{S}$, where $\mathbf{F}$ is $N\times
N$ matrix with the elements of
\begin{equation}
\label{eq:Fmatrix}
f_{k,n}=\frac{1}{\sqrt{N}}e^{j\frac{2\pi\alpha nk}{N}}
\end{equation}
according to \cite{IsaDar10}, SEFDM signal can be produced using
IFFT operation. It is assumed that at the receiver side the
transmitted signal, $x(t)$, is received with an Additive white
Gaussian Noise (AWGN); $r(t)=x(t)+n(t)$. To extract the sufficient
statistics from $r(t)$,  $N$ correlators are used at the receiver
side. The output of the $i$th correlator is calculated by
\begin{equation}
\label{eq:correlator}
R(i)=\int_0^T r(t)b_i^{*}(t)dt,\quad i=0,1,...,N-1
\end{equation}
In \cite{KanCho09} orthonormal function $b_i(t), i=0,1,...,N-1$
were used to prevent noise coloring and loss of information, but
in this paper , we do not limit ourselves to use only
orthonormal bases and like \cite{IsamIFFT}, the kernels
used for modulation at the transmitter are used as bases of the
receiver projection. As a result, we can benefit from a simple FFT
block instead of a correlator bank. If the output of the
correlator bank is shown by the vector
$\mathbf{R}=[R(0),R(1),...,R(N-1)]^T$, then
\begin{eqnarray}
\label{eq:Rmat}
\mathbf{R}&=&\mathbf{M}\mathbf{S}+\mathbf{N}
\end{eqnarray}
where the elements of the matrix $\mathbf{M}$ and the vector
$\mathbf{N}$ are defined by
\begin{eqnarray}
\label{eq:MNdef}
M_{ik}&=&\int_0^T b_i^*(t)e^{j2\pi k\Delta ft}dt\nonumber\\
N_{i}&=&\int_0^T n(t)b_i^*(t)dt
\end{eqnarray}
The main problem is to detect the vector of the transmitted
symbols $\mathbf{S}$ from the vector $\mathbf{R}$.
\section{iterative detection method}
\begin{figure}
\centering
\includegraphics[height=2in,width=3.5in]{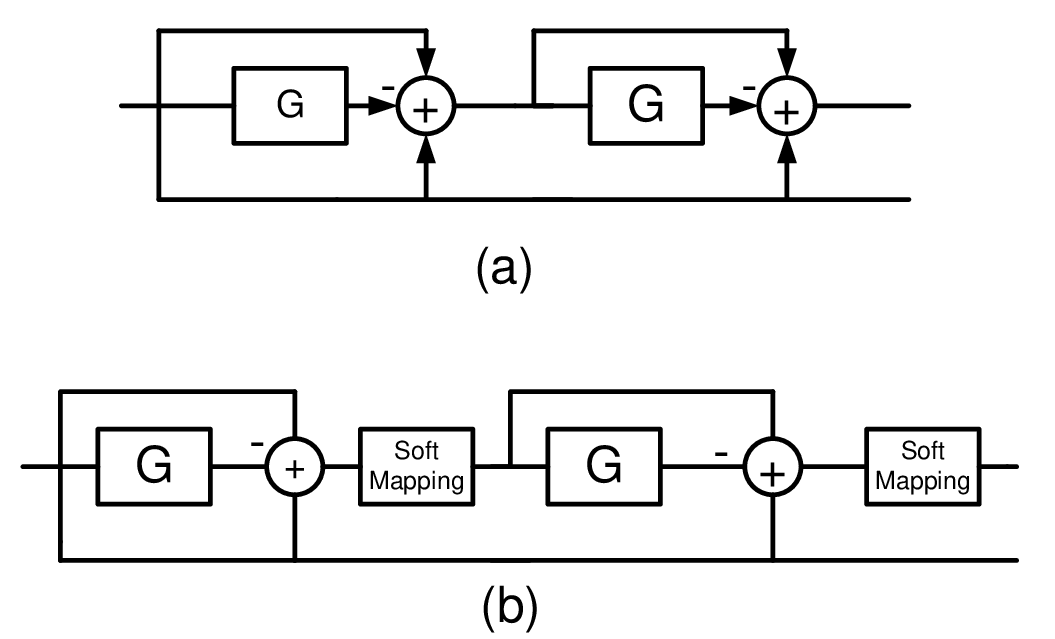}
\renewcommand{\figurename}{Fig.}
\caption{Block diagram of the iteration method for distortion
compensation a) Original method b) modified iterative method}
\label{fig:iter}
\end{figure}

A modified iterative method is proposed to be used for detection
of the transmitted symbols in SEFDM systems. The iterative
technique was first proposed by Marvasti \cite{Mar01} for the
cancellation of the interpolation distortion and then it was 
shown that this method could be used for the compensation of the
distortion caused by linear or nonlinear operations
\cite{MarKluwer99}, \cite{AminiMar07}. If a signal $\mathbf{S}$ is
distorted by the distortion operation $\mathbf{G}$ to produce
$\hat{\mathbf{S}}=\mathbf{G}\mathbf{S}$, then the signal
$\mathbf{S}$ can be reconstructed from the distorted
version, $\hat{\mathbf{S}}$, as follows (Fig.\ref{fig:iter}.a)
\begin{eqnarray}
\label{eq:Iter}
\mathbf{S}_n=\lambda \mathbf{S}_0+(1-\lambda G)\mathbf{S}_{n-1}.
\end{eqnarray}

where $\lambda$ is the relaxation parameter, $\mathbf{S}_n$ is
the output after $n$ iterations and $\mathbf{S}_0=\mathbf{G}\mathbf{S}$.
It has been shown in \cite{Mar01} that if the power of distortion,
$\Vert \mathbf{S}-\mathbf{G}\mathbf{S}\Vert_2$, is less than the
power of the signal, $\Vert \mathbf{S}\Vert_2$, then the iterative
method converge to the desired signal after an infinite number of
iterations. It has also been shown that if $\mathbf{G}$ is a nonlinear
distortion operation, the proper selection of the relaxation factor of
$\lambda$ can speed up the convergence.

In this paper, a modified version of this method is used to
compensate for the interference of the carriers in SEFDM systems.
In other words, FDM modulation and correlation bank in the SEFDM
systems are considered as a distortion function, $\mathbf{G}$, and
the iterative method is applied to compensate for the distortion,
i.e. $\mathbf{G}\mathbf{S}=\mathbf{M}\mathbf{S}$. It can be shown
that the value of Signal to Distortion Ratio (SDR) depends on the
number of carriers, $N$, and the frequency separation $\alpha$.
The closer to $1$ the value of $\alpha$ or the more the number of
carriers, the larger the SDR. Therefore, the iterative method converges faster
and to a better output.
\begin{figure}
\centering
\includegraphics[height=2.25in,width=2.75in]{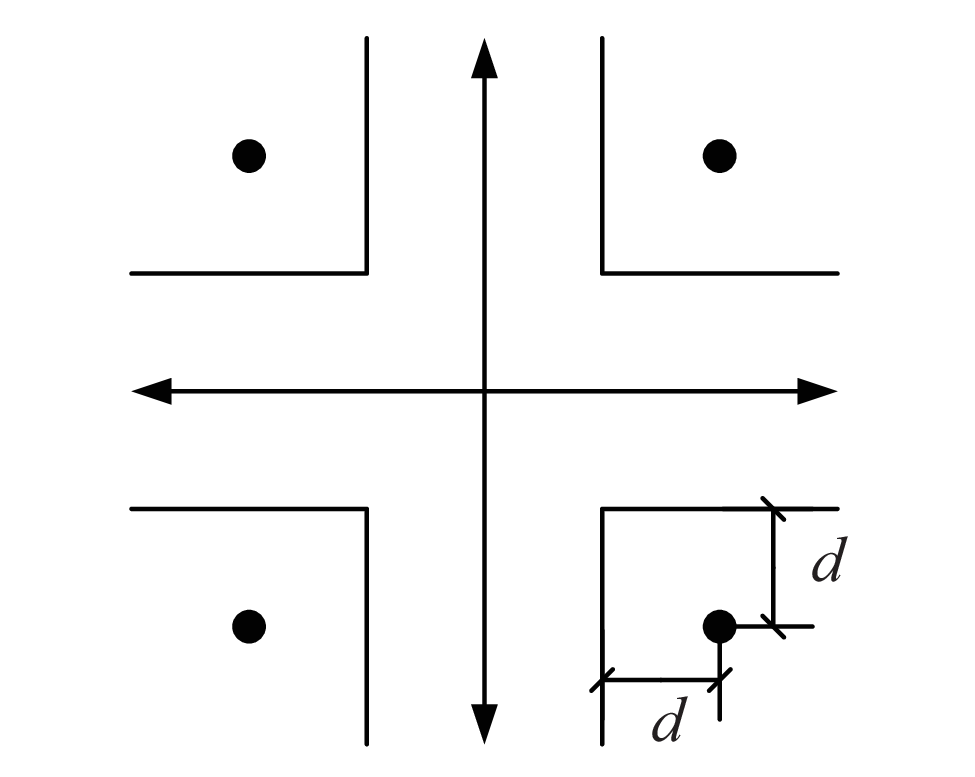}
\renewcommand{\figurename}{Fig.}
\caption{Soft mapping areas in 4-QAM modulation with parameter \emph{d}}
\label{fig:softmap}
\end{figure}

It is extra information that we know the elements of the
desired vector $\mathbf{S}$ belong to a constellation. Thus, in
the modified version of the iterative method at the output of each
iteration block, the points are mapped to the constellation points (as
shown in Fig.\ref{fig:iter}.b). This operation has the effect of
eliminating distortion and noise after each iteration, but it
may also have a disadvantage. If hard mapping is used all the
symbols are mapped to the nearest constellation point and it may
neutralize the effect of the iterations because it prevents the
movement of the symbols to the proper answer. Thus, a soft mapping
method is proposed instead of it. In this mapping
strategy, the symbols in the neighboring area of the constellation
points are mapped and the other ones are left unchanged. The
procedure of the mapping can be written by:
\begin{eqnarray}
\label{eq:softdec}
Q(S)=\left\{ \begin{array}{ll}
C_l & \textrm{if $ S\in A_l ,\quad l\in\{1,2,...,L\}$}\\
S & \textrm{if $ S\notin A_l$},
\end{array} \right.
\end{eqnarray}
where $C_1,C_2,...,C_L$ are the constellation points and $A_l$ is
the neighboring area around the constellation point $C_l$. For
4-QAM modulation the neighboring areas are defined by the
parameter $d$ as shown in Fig.\ref{fig:softmap}. The main
advantage of this mapping is that the points which are uncertainty and outside the defined areas
in Fig.\ref{fig:softmap}, are kept unchanged to be modified by
the iteration and the points that are very close to a
constellation point are decided. After each iteration, the number
of undecided symbols is reduced. Thus, the modified iteration
method can be described by:
\begin{eqnarray}
\label{eq:modifiediter}
\mathbf{S}_n&=&Q\bigg(\lambda\mathbf{S}_0+(1-\lambda \mathbf{G})\mathbf{S}_{n-1}\bigg)\nonumber\\
\mathbf{S}_0&=&\mathbf{F}^{-1}\mathbf{R}
\end{eqnarray}
Another method which is proposed in this paper is to adjust the
mapping areas in Fig.\ref{fig:softmap} dynamically. In this scheme the neighboring
areas $A_l,l=1,2,...,L$ are expanded after each iteration. As it
can be shown, a simple iteration without mapping has a very
moderate performance. Moreover, soft mapping has a better
performance than that of hard mapping. As a result, it should be
noted that soft mapping has a very influential role in our method
in order to achieve the desired BER.
\section{Analysis of Complexity}
The main advantage of the proposed method, in comparison to SD and
ML methods introduced in \cite{RodDar03}, is its low complexity. The
complexity of these methods can be evaluated by number of Real
Additions (RA) and Real Multiplications (RM). In this section, the
order of complexity according to this criterion is computed for ML,
SD and the proposed  iterative methods.\\

In ML method, the following metric must be calculated for $L^N$
different possible vectors $\mathbf{S}$:
\begin{equation}
\label{eq:MLmetric}
\Vert\mathbf{R}-\mathbf{M}\mathbf{S}\Vert_2
\end{equation}
The computation of $\mathbf{M}\mathbf{S}$ for each vector
$\mathbf{S}$ consists of an $N/\alpha$ points IFFT and an
$N/\alpha$ points FFT which needs $2N/\alpha\log_2(N/\alpha)$ RMs
and $6N/\alpha\log_2(N/\alpha)$ RAs \cite{OppSch}. Moreover, $4N$
RAs and $2N$ RMS are needed to obtain L2 norm of the error vector
in (\ref{eq:MLmetric}). Consequently, $2L^NN/\alpha
(3\log_2(N/\alpha)+2\alpha)$ RAs and $2L^NN/\alpha
(\log_2(N/\alpha)+\alpha)$ RMs should be performed for every data block.\\

In SD method, the metric (\ref{eq:MLmetric}) is calculated for
some possible vectors $\mathbf{S}$ which are in a sphere of radius
$d$ defined by $\Vert\mathbf{R}-\mathbf{M}\mathbf{S}\Vert_2\leq
d^2$. In this method, the complexity depends on the number of
nodes in this sphere. This number is a random variable which
depends on $SNR$ and the radius $d$. Thus, the expected
complexity of SD method can be calculated. In \cite{JalOtt05}, the
expected complexity of SD method has been derived in a general
form. It has been seen that, the expected number of operations is
in order of $L^{\gamma N}$, where $0<\gamma<1$ is a small factor
that is dependent on SNR and $d$. However, for large values of
SNR, the factor $\gamma<<1$, and therefore, $L^{\gamma N}$ is
close to $1$ when $N$ is small. This means that for large values
of  SNR and small $N$, the complexity is dominated by polynomial
terms, which is consistent with the results of \cite{HasVik02},
but in low SNR regimes, it is NP hard. As a result, the numbers
of the required RAs and RMs in SD method are like those of the ML
method except that $L^N$ must be replaced
by $L^{\gamma N}$.\\

In the proposed method, at each iteration the operation
$\mathbf{G}$ must be performed which consists of an
$N$-points IFFT and an $N$-points FFT. After
calculation of the term $\mathbf{G}\mathbf{S}_{n-1}$, the other
operations in (\ref{eq:modifiediter}) need $4N$ RAs and $2N$ RMs.
For soft mapping, which is performed at the end of every iteration
block, the symbols must be compared with all of the constellation
points. If we assume that the complexity of this comparison is
equivalent to the complexity of one RA, then the complexity of
soft mapping is equivalent to that of $LN$ RAs. As a result, in
every iteration, $N(L+4+6/\alpha\log_2(N/\alpha))$ RAs and 
$N(2/\alpha\log_2(N/\alpha)+2)$ RMs are performed.
\section{Simulation Results}
\begin{figure}
\centering
\includegraphics[height=2.5in,width=3.75in]{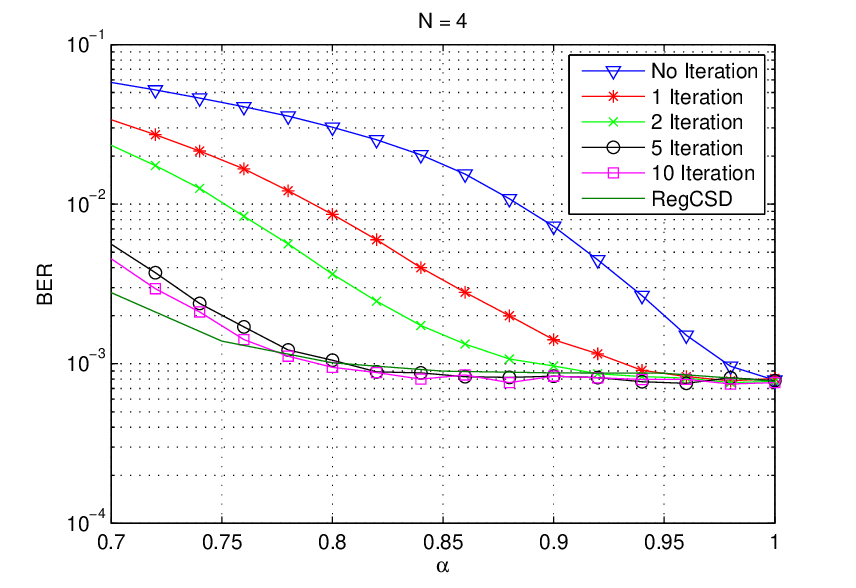}
\renewcommand{\figurename}{Fig.}
\caption{BER versus $\alpha$ for SEFDM system with $N=4$ carriers
and different detection methods: SD and the proposed method with
$1$, $2$, $5$ and $10$ iterations.} \label{fig:sim1}
\end{figure}

\begin{figure}
\centering
\includegraphics[height=2.5in,width=3.75in]{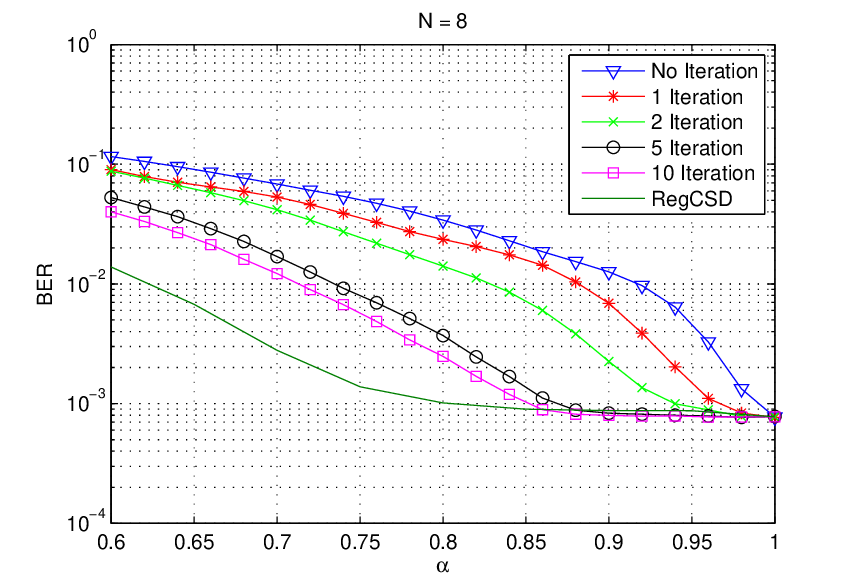}
\renewcommand{\figurename}{Fig.}
\caption{BER versus $\alpha$ for SEFDM system with $N=8$ carriers
and different detection methods: SD and the proposed method with
$1$, $2$, $5$ and $10$ iterations.} \label{fig:sim2}
\end{figure}

\begin{figure}
\centering
\includegraphics[height=2.5in,width=3.75in]{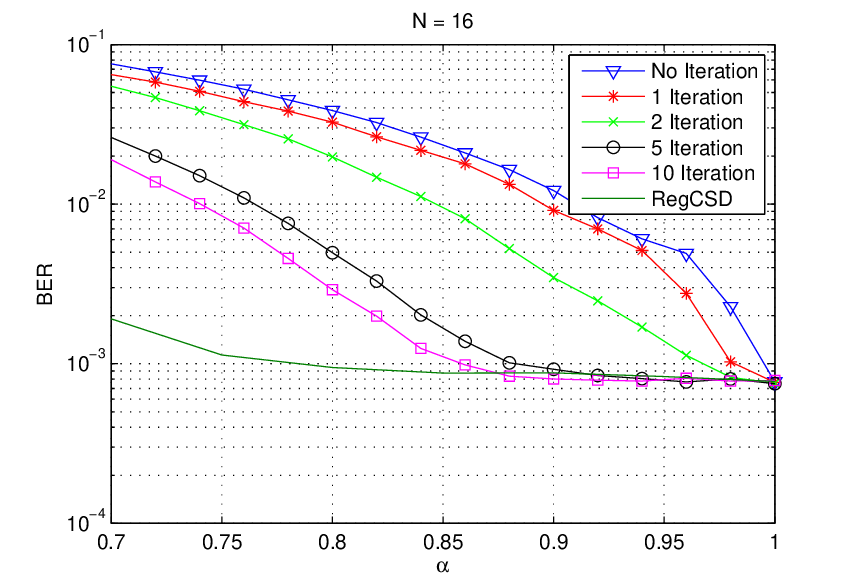}
\renewcommand{\figurename}{Fig.}
\caption{BER versus $\alpha$ for SEFDM system with $N=16$ carriers
and different detection methods: SD and the proposed method with
$1$, $2$, $5$ and $10$ iterations.} \label{fig:sim3}
\end{figure}

To evaluate the performance of the proposed iterative method,
SEFDM systems with 4-QAM modulation and different numbers of
carriers, $N$, and different values of the parameter $\alpha$ have
been simulated. Regularized Complex SD (RegCSD) is used as a near optimal 
method for comparison. Figures \ref{fig:sim1}, \ref{fig:sim2} and
\ref{fig:sim3} show the Bit Error Rate (BER) versus $\alpha$ for
$N=4$, $N=8$ and $N=16$, respectively. The SNR is fixed at $10dB$
in these figures. In this simulation, the parameter $d$ for soft
mapping starts from $1$ at the first iteration and is reduced
linearly to $0$. As it can be seen, for $N=4$ the performance of the
proposed iterative method with $5$ iterations is very close to
that of near optimal SD method, while (as shown in table I) the
iterative method has a lower complexity than that of the SD
method. It is clear that the $BER$ is increased when $\alpha$ is
decreased. This is more evident when the number of
subcarriers is increased form $N=4$ to $N=16$. The most important 
result of these figures is that for $\alpha>0.85$ performance of iterative 
method is identical to SD one.

\begin{figure}
\centering
\includegraphics[height=2.5in,width=3.75in]{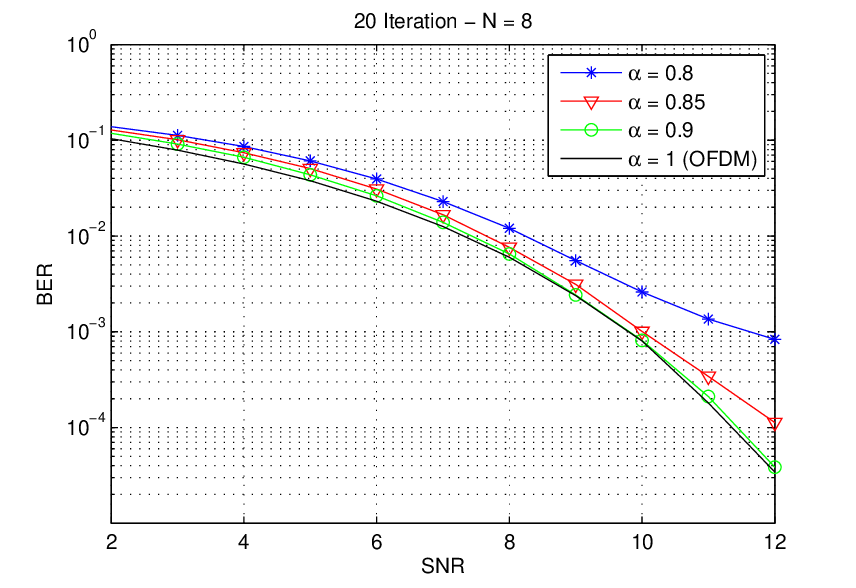}
\renewcommand{\figurename}{Fig.}
\caption{BER versus SNR for SEFDM system with $N=8$ carriers and
different detection methods: SD and the proposed method with $20$
iterations.} \label{fig:sim4}
\end{figure}

\begin{figure}
\centering
\includegraphics[height=2.5in,width=3.75in]{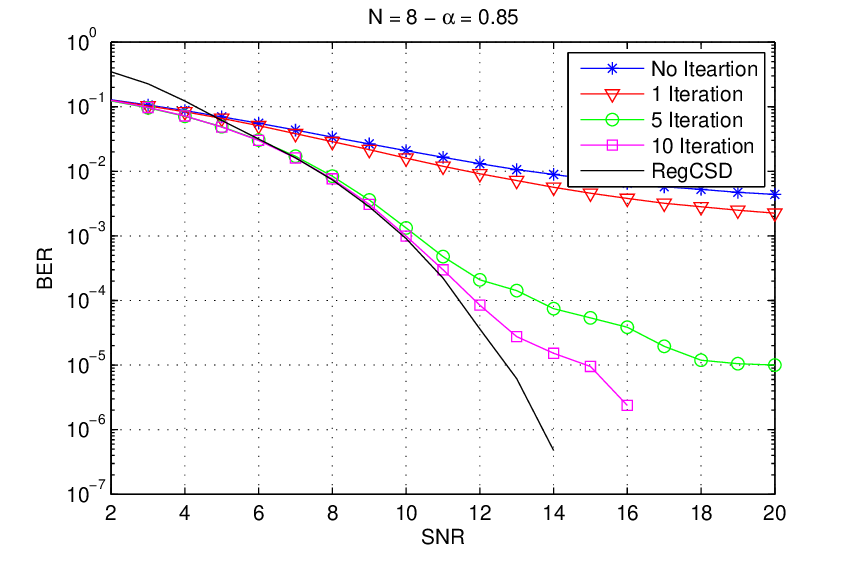}
\renewcommand{\figurename}{Fig.}
\caption{BER versus SNR for SEFDM system with $N=8$ carriers and
different detection methods: SD and the proposed method.}
\label{fig:sim5}
\end{figure}

Figure \ref{fig:sim4} shows BER versus SNR for different values
of $\alpha$ and $N=8$ for proposed method after $10$ iterations. 
In this figure the BER performance of OFDM system ($\alpha=1$) 
has also been shown. As it is clear
from this figure, the performance degradations in comparison to
OFDM systems are $1dB$, $2.2dB$ and $5dB$ for $\alpha=0.9$, $0.85$
and $0.8$, respectively. Thus, the performance of the proposed
method is acceptable for the overlapping factors of less than
$15\%$. Figure $\ref{fig:sim5}$ shows the BER performance of the
proposed method for $N=8$ and $\alpha=0.85$ versus SNR. It is
obvious that the proposed method effectively reduce the BER at
the receiver side and can compete SD method, specially in low SNR regime.

Table I compares the ratio of simulation time for the SD method
over a single iteration of our proposed method. Since our method
converges after $10$ iteration, it is apparent that we gain a
great advantage in term of complexity by our method.\\

\begin{table}{}
\centering \caption{Simulation Time Ratio $\gamma$ of SD over a single iteration}
\resizebox{8cm}{!} {
\begin{tabular}{|c|c|c|c|}
\hline
$\gamma$ & $\alpha = 0.8$ & $\alpha = 0.85$ & $\alpha = 0.9$ \\
\hline
$N = 4$  & 170  & 160  & 155 \\
\hline
$N = 8$  & 710  & 645  & 640 \\
\hline
$N = 16$ & 4800 & 4250 & 4160 \\
\hline
\end{tabular}
}
 \label{tab:sim1}
\end{table}

\section{Conclusion}
In this paper, a modified version of an iterative method was
proposed for interference reduction in SE-FDM systems. In this
method, after each iteration the output is mapped to the
constellation points using a soft mapping operation. The soft
mapping operation is dynamically changed to the hard mapping after
each iteration by changing the decision areas adaptively. The
performance of the proposed method was compared with the near
optimal SD method. Simulations show that this method has
approximately the same performance as that of SD method with much
lower complexity when the overlapping is less than $15\%$.

\ifCLASSOPTIONcaptionsoff
  \newpage
\fi
\bibliographystyle{IEEEtran}

\begin{thebibliography}{30}
\bibitem{SamTal02} H. Sampath, S. Talwar, J. Tellado, V. Erceg, and A. Paulraj,
"A Fourth generation MIMO-OFDM Broadband Wireless System: Design, Performance,
and Field Trial Results," \emph{IEEE Communications Magazine}, vol. 40, no. 9,
pp. 143-149, Sep. 2002.

\bibitem{RapAnn02} T. S. Rappaport, A. Annamalai, R. M. Buehrer, and W. H. Tranter,
"Wireless Communications: Past Events and a Future Perspective,"
\emph{IEEE Communications Magazine}, vol. 40, no. 5, pp. 148-161,
May 2002.

\bibitem{RodDar02} M. Rodrigues and I. Darwazeh,
"Fast OFDM: A proposal for doubling the data rate of OFDM
schemes," \emph{in Proceedings of the International Conference on
Telecommunications}, Beijing, China, June 2002, pp. 484-487.

\bibitem{Xio03} F. Xiong,
"M-ary amplitude shift keying OFDM system," \emph{IEEE
Transactions on Communications}, vol. 51, no. 10, pp. 1638-1642,
Oct. 2003.

\bibitem{RodDar03} M. Rodrigues and I. Darwazeh,
"A spectrally efficient frequency division multiplexing based
communication system," \emph{in 8th International OFDMWorkshop},
Hamburg, Germany, Sept. 2003, pp. 70-74.

\bibitem{HamTac04} M. Hamamura and S. Tachikawa,
"Bandwidth efficiency improvement for multi-carrier systems,"
\emph{15th IEEE International Symposium on Personal, Indoor and
Mobile Radio Communications, PIMRC 2004}, vol. 1, p. 48-52 Vol.1,
Sept. 2004.

\bibitem{RusAnd05}F. Rusek and J. B. Anderson,
"The Two Dimensional Mazo Limit," \emph{in International Symposium
of Information Theory, ISIT 2005},vol. 57, pp. 970-974, 2005.

\bibitem{MazLan88} J. Mazo and H. Landau,
"On the minimum distance problem for fasterthan-Nyquist
signaling," \emph{IEEE Transactions on Information Theory}, vol.
34, no. 6, pp. 1420-1427, Nov 1988.

\bibitem{Maz75}J. E. Mazo,
"Faster than Nyquist Signalling," \emph{Bell Systems Technical
Journal}, vol. 54, pp. 429-458, Oct 1975.

\bibitem{AhmDar09} S. I. A. Ahmed and I. Darwazeh,
"IDFT Based Transmitters for Spectrally Efficient FDM System,"
\emph{in London Communication Symposium}, Sep 2009.

\bibitem{IsaDar10} S. Isam and I. Darwazeh,
"Simple DSP-IDFT Techniques for Generating Spectrally Efficient
FDM Signals," \emph{in IEEE, IET International Symposium on
Communication Systems, Networks and Digital Signal Processing},
pp. 20 - 24, Jul 2010.

\bibitem{IsaChor08} I. Kanaras, A. Chorti, M. Rodrigues, and I. Darwazeh,
"A combined MMSE-ML detection for a spectrally efficient non
orthogonal FDM signal," \emph{5th International Conference on
Broadband Communications, Networks and Systems}, pp. 421 − 425,
Sept. 2008. 3.

\bibitem{IsaKan2011} S. Isam, I. Kanaras, I. Darwazeh,
"A Truncated SVD approach for fixed complexity spectrally
efficient FDM receivers," \emph{Wireless Communications and
Networking Conference (WCNC), 2011 IEEE}, pp.1584-1589, 28-31
March 2011

\bibitem{KanChoRod09} I. Kanaras, A. Chorti, M. Rodrigues, and I. Darwazeh,
"Spectrally Efficient FDM Signals: Bandwidth Gain at the Expense
of Receiver Complexity," \emph{in Proceedings of the International
Conference On Communications}, pp. 1-6, 2009.

\bibitem{KanChoRodJapan09} I. Kanaras, A. Chorti, M. Rodrigues, and I. Darwazeh,
"Investigation of a Semidefinite Programming detection for a spectrally efficient
FDM system," \emph{20th Personal, Indoor and Mobile Radio Communications Conference
2009, IEEE PIMRC'09, Japan, Tokyo}, September 2009, pp. 2827 − 2832.

\bibitem{KanCho09} I. Kanaras, A. Chorti, M. Rodrigues, and I. Darwazeh,
"A new quasioptimal detection algorithm for a non orthogonal
Spectrally Efficient FDM," \emph{in Proc. 9th Int. Symp.
Communications and Information Technology ISCIT 2009}, pp.
460-465, 2009.

\bibitem{IsaDar11} S. Isam, I. Darwazeh,
"Design and Performance Assessment of Fixed Complexity Spectrally
Efficient FDM Receivers," \emph{73rd IEEE Vehicular Technology
Conference (VTC Spring) 2011}, pp.1-5, 15-18 May 2011

\bibitem{Precoding} S. Isam, I. Darwazeh,
"Precoded Spectrally Efficient FDM System," \emph{in 21th
Personal, Indoor and Mobile Radio Communications Symposium (IEEE
PIMRC'10)}, pp. 99-104, Sep 2010.

\bibitem{Mar01} F. Marvasti,
"An iterative method to compensate for the interpolation
distortion," \emph{IEEE Trans. ASSP}, Vol. 3, No. 1, 1617-1621,
1989.

\bibitem{IsamIFFT} S. Isam, I. Darwazeh,
"Design and Performance Analysis of Enhanced Receivers for
Spectrally Efficient FDM System," \emph{in London Communication
Symposium (LCS'11)}, Sep 2011

\bibitem{MarKluwer99} F. Marvasti,
"Nonuniform Sampling: Theory and Practice", Kluwer Academic/Plenum
Publishers, 2001. Trans. Med. Imag., Vol. 18, No. 11, Nov. 1999.

\bibitem{AminiMar07} A. Amini, F. Marvasti,
"Reconstruction of multiband signals from non-invertible uniform
and periodic nonuniform samples using an iterative method,"
\emph{in Proceedings of SampTa 2007}, Thessaloniki, Greece, June
2007.


\bibitem{OppSch} A.V. Oppenheim, R.W. Schafer, and J.R. Buck,
"Discrete-time signal processing", third edition. Prentice-Hall,
Inc.: Upper Saddle River, NJ, 2009.

\bibitem{JalOtt05} J. Jalden, B. Ottersten,
"On the complexity of sphere decoding in digital communications,"
\emph{IEEE Trans. on Signal Processing}, vol.53, no.4, pp. 1474-
1484, April 2005.

\bibitem{HasVik02}B. Hassibi, H. Vikalo,
"On the expected complexity of integer least-squares problems,"
\emph{IEEE International Conference on Acoustics, Speech, and
Signal Processing (ICASSP 2002)}, vol.2, no. , pp.II-1497-II-1500,
13-17 May 2002.

\end{thebibliography}

\end{document}